\documentclass[10pt,conference]{IEEEtran}
\usepackage{cite}
\usepackage{amsmath,amssymb,amsfonts}
\usepackage{algorithmic}
\usepackage{graphicx}
\usepackage{textcomp}
\usepackage{xcolor}
\usepackage{acro}
\usepackage{listings}
\graphicspath{{./figures/}}

\def\BibTeX{{\rm B\kern-.05em{\sc i\kern-.025em b}\kern-.08em
    T\kern-.1667em\lower.7ex\hbox{E}\kern-.125emX}}

\DeclareAcronym{oss}{
    short = OSS,
    long = open source software 
}

\DeclareAcronym{scm}{
    short = SCM,
    long = source code management 
}

\DeclareAcronym{pr}{
    short = PR,
    long = pull request,
    short-plural = s,
    long-plural = s,
}

\DeclareAcronym{dag}{
    short = DAG,
    long = directed acyclic graph,
    short-plural = s,
    long-plural = s,
}

\DeclareAcronym{tkg}{
    short = TKG,
    long = temporal knowledge graph,
    short-plural = s,
    long-plural = s,
}
\DeclareAcronym{kg}{
    short = KG,
    long = knowledge graph,
    short-plural = s,
    long-plural = s,
}

\DeclareAcronym{msr}{
    short = MSR,
    long = Mining Software Repositories,
}

\title{Git2KG\@: Mining socio-technical relationships in GitHub repositories as temporal knowledge graphs}

\author{\IEEEauthorblockN{Lennart Hessenauer}
    \IEEEauthorblockA{\textit{CAIDAS} \\
        \textit{Julius-Maximilians-Universität}\\
        Würzburg, Germany \\
        lennart.hessenauer@uni-wuerzburg.de}
    \and
    \IEEEauthorblockN{Lisi Qarkaxhija}
    \IEEEauthorblockA{\textit{CAIDAS} \\
        \textit{Julius-Maximilians-Universität}\\
        Würzburg, Germany \\
        lisi.qarkaxhija@uni-wuerzburg.de}
    \and
    \IEEEauthorblockN{Ingo Scholtes}
    \IEEEauthorblockA{\textit{CAIDAS} \\
        \textit{Julius-Maximilians-Universität}\\
        Würzburg, Germany \\
        ingo.scholtes@uni-wuerzburg.de}
}

\begin{document}

\bstctlcite{IEEEexample:BSTcontrol}
\maketitle
\begin{abstract}
The widespread use of Git and GitHub in \ac{oss} projects has produced large volumes of data on collaborative software development. 
Existing approaches to mine those data often use graph models to (i) capture communication between developers in the issue tracker, or (ii) represent developer-developer co-authorship or developer-file code ownership networks from the editing history of code artifacts. 
Integrating these data in a single \ac{kg} is complicated by the need to disambiguate authors and match entities from multiple sources. 
Moreover, most works using network analysis have taken a time-aggregated perspective that neglects temporal dynamics of collaborations.

To address these issues, we present \texttt{Git2KG}, an extensible Python library that mines \acp{tkg} from GitHub repositories. 
\texttt{Git2KG} extracts fine-grained co-editing relations and code ownership from the Git log. It further disambiguates developer identities to align the resulting graphs with management and communication networks extracted from GitHub issue trackers. 
\texttt{Git2KG} supports temporal analysis by slicing development events into fixed-size, non-overlapping time windows, yielding a series of knowledge graph snapshots per repository. 
It further supports computing node features and labels for downstream machine learning tasks. 
Snapshots can then be exported to PyTorch Geometric, PathPyG, and NetworkX for temporal network analysis, visualization, and graph learning.

\end{abstract}

\section{Introduction}
\ac{oss} has become increasingly popular over the last decades and forms the backbone of our digital society. 
Such software is often developed by a large and volatile community of developers, which introduces challenges in the management of source code contributions and communication between developers and users. 
Git and GitHub, specifically its issue tracker, address those challenges: Git facilitates the management of source code, while the GitHub issue tracker provides a platform for communication. 
The widespread usage of those systems introduces large volumes of data about the software development process that are logged in the respective systems. 
These data have attracted interest in the research field \ac{msr}~\cite{DBLP:conf/sigsoft/SpadiniAB18,DBLP:conf/msr/BirdRBHGD09,DBLP:conf/msr/KalliamvakouGBSGD14}.

One popular way to model these data is to encode the complex relations between contributors, issues, and source code as graphs~\cite{DBLP:conf/cscw/CataldoH08,DBLP:conf/msr/BirdGDGS06,DBLP:conf/oss/RobertsaHS06,DBLP:conf/indiaSE/SurekaGR11,DBLP:conf/icse/ZanettiSTS13a,DBLP:conf/icse/ZanettiSTS13,DBLP:conf/oss/HowisonIC06,DBLP:conf/msr/HuangL05, DBLP:conf/sigsoft/MeneelyWSO08,DBLP:journals/corr/abs-2008-04884,DBLP:conf/msr/MacLeanK13,DBLP:conf/icse/GoteMSS22}. 
However, most methods either focus on the relationships between developers and code, co-authorship dependencies between developers, or links that model communication patterns of contributors. 
The \ac{msr} community lacks a reusable graph mining tool which captures all of these relationships -- spanning from fine-grained source code edits over commit and code ownership, to communication between users on and with the GitHub issue tracker -- in a single comprehensive \acf{kg}. 

To fill this gap, we present \texttt{Git2KG}, a Python library which supports the mining of \acp{tkg} with arbitrary time resolution from repositories on GitHub. 
\texttt{Git2KG} leverages data from the Git log and the GitHub issue tracker, mining detailed knowledge graph models which represent different types of relations in \ac{oss} development. 
To connect data from the two sources, we align authors and committers in Git and GitHub user accounts, using identity disambiguation. 
The \acp{kg} mined by \texttt{Git2KG} are suitable for network analysis, visualization, and machine learning. 
For the latter, \texttt{Git2KG} provides extendable engines for node feature and label computation. 
    We provide the tool through GitHub\footnote{https://github.com/lennart-h/git2kg}.

The remainder of this work is structured as follows: In Section~\ref{sec:rel-work}, we present existing graph models and graph mining tools modelling the software development process.
Section~\ref{sec:lib-design} presents the schema of the \acp{tkg} mined by \texttt{Git2KG}, the data loading pipeline, entity disambiguation, as well as the extendable feature and label computation and export modules. 
In Section~\ref{sec:evaluation}, we evaluate the mining performance and actor alignment of our library. 
Finally, we conclude the work in Section~\ref{sec:conclusion} by showing the limitations of our approach and pointing toward further research. 

\section{Related Work}\label{sec:rel-work} 
A number of works in the context of repository mining use graphs or networks to model events from software engineering. 
Developer communication has been studied by applying social network analysis to data from IRC channels~\cite{DBLP:conf/cscw/CataldoH08}, mailing lists~\cite{DBLP:conf/msr/BirdGDGS06,DBLP:conf/oss/RobertsaHS06}, and bug trackers~\cite{DBLP:conf/indiaSE/SurekaGR11,DBLP:conf/icse/ZanettiSTS13a,DBLP:conf/icse/ZanettiSTS13,DBLP:conf/oss/HowisonIC06,DBLP:conf/icse/WolfSDN09}.
Co-authorship networks instead link developers who edited the same file~\cite{DBLP:conf/msr/HuangL05, DBLP:conf/sigsoft/MeneelyWSO08}.
MacLean et al.~\cite{DBLP:conf/msr/MacLeanK13} enhanced such models to \acp{kg} connecting authors, revisions, and files based on code authorship and file modifications. 
The popularity of GitHub motivated researchers to model data from the platform as graphs, including reviewing behavior~\cite{ZOLLER2020120291}, collaboration~\cite{DBLP:journals/sqj/OliveiraMBBHM23}, comment networks~\cite{DBLP:journals/infsof/YuWYW16}, or project follower relationships~\cite{DBLP:journals/corr/abs-2502-00058, DBLP:journals/corr/YanW17,DBLP:journals/eswa/HuWRC18}.

Beyond such single datasets, several tools to mine graphs from repositories exist.
GraphRepo~\cite{DBLP:journals/corr/abs-2008-04884} extracts multi-relational \acp{kg} connecting developers, commits, branches, files, and methods from the Git development history. 
TNM~\cite{DBLP:conf/msr/SviridovEK21} is a Kotlin tool which mines 
``people networks'' and ``software networks'' from Git and connects the two subgraphs with relations between people and software components. 
Kaiaulu~\cite{DBLP:conf/ecsa/ParadisK21a} mines socio-technical networks combining email communication and coarse-grained file authorship. 

As developers editing the same file may change unrelated code regions in the file, git2net~\cite{DBLP:conf/msr/GoteSS19} defines co-editing relations on the line level instead of file-based co-authorship as the notion for collaboration. 
It builds temporal co-editing networks using the Git log of a repository, which have been used to study productivity~\cite{DBLP:conf/icse/GoteMSS22}.

To the best of our knowledge, no network model or mining tool connects fine-grained co-edits with code ownership and communication on GitHub. \texttt{Git2KG} fills that gap.
\begin{figure}
    \centering
    \includegraphics[width=.9\linewidth]{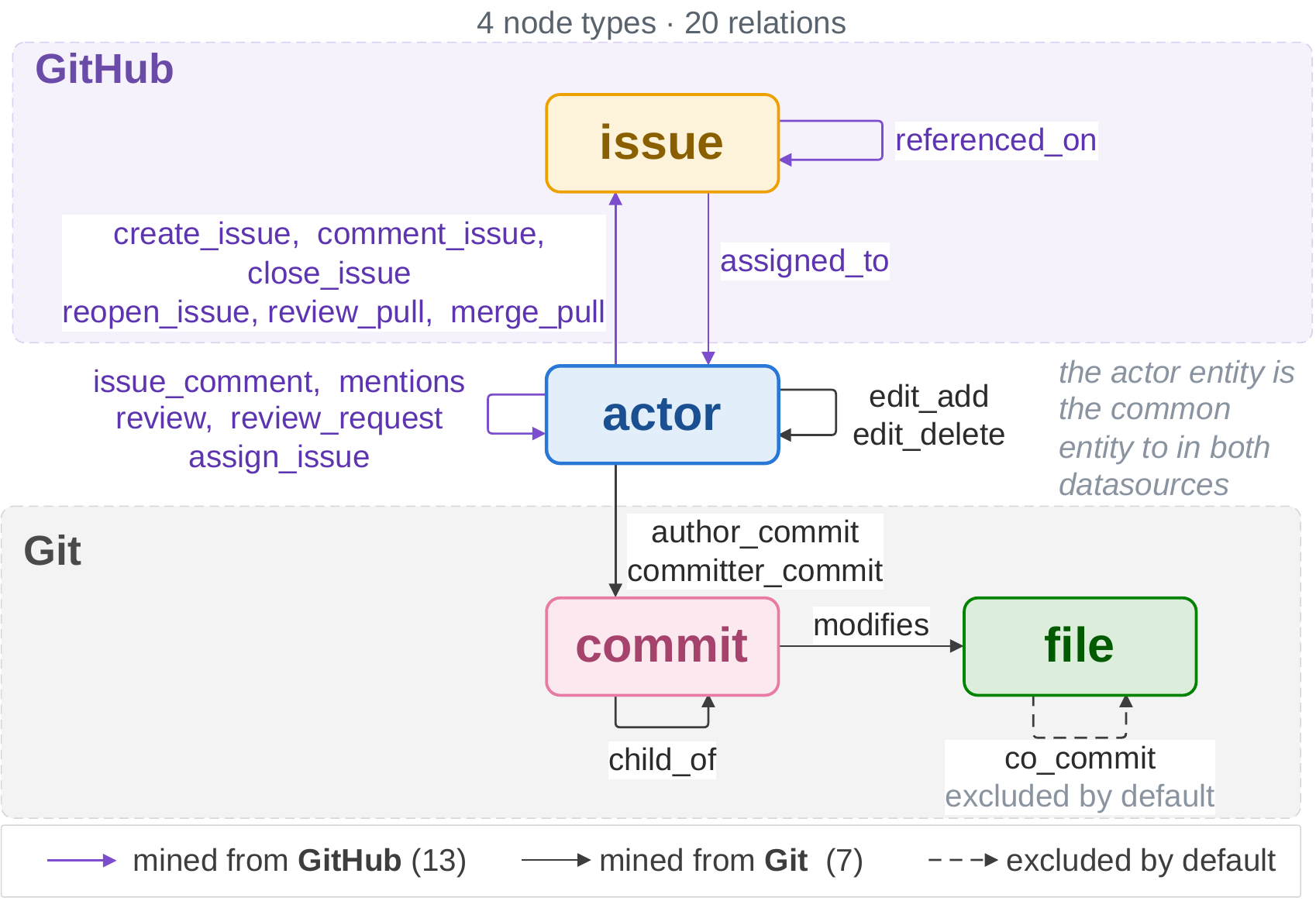}
    \caption{Schema of \texttt{Git2KG} knowledge graphs. The color encodes the data source. The actor entity appears in both sources as the aligning entity.}\label{fig:schema}
\end{figure}

\section{Mining Pipeline and Library Design}\label{sec:lib-design}
In this section, we discuss the design of \texttt{Git2KG} and how it mines \acp{tkg} from GitHub repositories. 
We explain the \ac{kg} schema, data loading in \texttt{Git2KG}, and identity disambiguation. 
Next, we show how we use time slicing to gain a sequence of \acp{kg}, before finishing the pipeline with feature and label computation and exports to other data formats. 

\subsection{Knowledge Graph Schema}
\texttt{Git2KG} mines \acp{kg} with four entity types: actors, commits, files, and issues, where \acp{pr} are treated as a special kind of issue. 
Fig.~\ref{fig:schema} shows the full schema. 
Commits and files are mined exclusively from Git, and issues exclusively from GitHub. 
Actors appear in both sources, as commit authors and committers in Git and as issue tracker users on GitHub, and therefore serve as the entity that aligns the two subgraphs. 
\texttt{Git2KG} extracts $20$ typed relations between these entities, seven from the Git history and $13$ from the GitHub issue tracker. 
Each relation corresponds to a timestamped development event, which allows us to slice the graph into snapshots. 

From Git, we extract co-editing relations between actors following git2net~\cite{DBLP:conf/msr/GoteSS19}: an \texttt{edit\_add} (\texttt{edit\_delete}) edge links an actor who adds (deletes) code to the author who last edited the affected line. 
The \texttt{author\_commit} and \texttt{committer\_commit} relations declare the commit ownership of actors, \texttt{child\_of} encodes the relation between a parent and a child commit, and the \texttt{modifies} relation links commits to the files in their change set. 
\texttt{Git2KG} also supports the extraction of the \texttt{co\_commit} relation, which links files that are changed in the same commit, but excludes the extraction of this relation by default as it is a sole projection of the \texttt{modifies} relation and produces only clique subgraphs with no additional information encoded. 

From the GitHub issue tracker, we extract the \texttt{create\_issue}, \texttt{comment\_issue}, \texttt{close\_issue}, \texttt{reopen\_issue}, \texttt{review\_pull}, and \texttt{merge\_pull} relations from actors to issues directly from the issue events and the \texttt{assigned\_to} relation from issues to actors.  
The \texttt{cross-referenced} event of the issue tracker yields the \texttt{referenced\_on} relation between two issues if one is referenced on the other. 
Following~\cite{DBLP:conf/oss/HowisonIC06}, the \texttt{issue\_comment} relation links two actors if the first commented right after the second on the same issue. 
The \texttt{mentions} relation encodes whenever an actor uses the @-mention feature to address a comment directly to another actor. 
The relations \texttt{review}, \texttt{review\_request} and \texttt{assign\_issue} encode the actor-to-actor interactions while solving an issue or \ac{pr}.

\subsection{Data Loading}
\texttt{Git2KG} loads data from both Git and GitHub using the \texttt{GitLoader}, \texttt{GitHubLoader}, and \texttt{GitHubGraphQLLoader} classes, which are proxied by the \texttt{Dataloader} class.
It unifies the interface for data loading and streams the loaded data points to an implementation of the \texttt{PersistenceManager}. 
Implementations of that interface save and retrieve data points using a persistence backend, and enable incremental restarts of data loading.
We provide a simple implementation of the \texttt{PersistenceManager} based on CSV files.
Users of the package can, however, provide their own implementations of the interface, e.g., to integrate the data loading into existing ecosystems.

\texttt{GitLoader} loads data from Git by traversing the commit history in a repository from the first to the last commit using PyDriller~\cite{DBLP:conf/sigsoft/SpadiniAB18}.
Doing so, it extracts data points corresponding to commits, line edits, and modified files.
Commits and modified files are directly read from the commit log, but co-edits are only implicitly encoded in the version history of Git.
For every file in a commit, we use PyDriller's \texttt{ModifiedFile} to get a list of modifications in the file, apply \texttt{blame} on the previous version (parent commit) of the file, and merge the two lists. 
This yields pairs of previous authors and edits per modified line.

We conduct data loading from Git on an offline clone of the repository to not be constrained by rate limits or third-party API availability. 
The loading of those data points is, however, still relatively expensive in computation. 
To minimize the loading time for the data points, we split the commit set of a repository into chunks of $20$ commits and process them in parallel. 

\texttt{GitHubLoader} and \texttt{GitHubGraphQLLoader} are interchangeable loaders for issues, issue comments, reviews, events, and issue timeline events, using the GitHub REST API and GraphQL API, respectively.
Both loaders require a GitHub access token to gain access to their corresponding API\@. 
GraphQL retrieves related data points (comments, reviews, events, etc.) together with their issue, which reduces rate-limit overhead, at the cost of persisting them only after the download completes instead of incrementally during the download. 
It also provides HTML bodies, which enables us to use hovercard markup to extract @-mentions instead of relying on pattern matching. 
This avoids false positives from code snippets like Java annotations or Python decorators. 
We therefore use GraphQL by default.
\subsection{Entity Alignment and Disambiguation}\label{sub:alignment}
As we encode the development process across long time ranges with \texttt{Git2KG}, it is crucial that the identifiers of the entities are stable over time.
This is unproblematic for issues and commits, which have project-global immutable identifiers (issue number and hash).

Files are identified by their path, which changes when a file is moved or renamed. 
We therefore assign each file a numerical ID when it is added and follow it through renames, using the modification types recorded in the commit history. 
This yields a stable ID for every modified file in every commit.

Actors are the only entity present in both sources, so aligning Git identities with GitHub accounts connects the two subgraphs. 
While users on GitHub carry a unique numeric identifier, Git identifies authors and committers only by name and email, and one person may use several of each. 
We therefore run gambit~\cite{DBLP:conf/msr/GoteZ21}, a disambiguation tool which matches name–email pairs through string-similarity and exact-match rules, on all identities from Git and GitHub.
GitHub users often keep their email addresses private, so gambit cannot match them if their names also differ. 
For these cases, we use GitHub’s commit API, which links commits to the account whose verified email matches the commit email, and merge the corresponding actors.

\subsection{Time Slicing}
As all events in Git and GitHub are timestamped, the relations mentioned above carry temporal information.  
To encode this temporal information in a \ac{tkg}, we slice the loaded data into non-overlapping time windows of equal size. 
The granularity of the time slices is fully configurable by the user. 
One time slice contains all data points whose timestamps are equal to or greater than the lower border of the time slice and strictly less than the upper border of the time slice. 
We then extract the edges between the nodes for each time slice based on the corresponding data points, ensuring temporal integrity of extracted edges.
Extracting one \ac{kg} per time slice yields a series of time-aggregated \ac{kg} snapshots per repository, i.e., a \ac{tkg}.

\subsection{Features, Labels, and Exports}
Apart from creating edges between entities, we can also use the extracted data to compute features and labels of entities for downstream machine learning tasks. 
The computation of features is handled by entity-type-specific implementations of the \texttt{FeatureExtractor} class, which can be seamlessly extended by users.
On \ac{kg} snapshot creation, all provided features are computed and then compiled into a feature matrix.
Features change as more information becomes available over time, while not being instantaneous like edges, but describing an entity throughout its lifetime. 
As a consequence, we provide the full historical data to the feature extractors, by not applying the lower bound constraint to time slicing. 
Node labels are computed analogously using implementations of the \texttt{LabelExtractor} class.
As labels might be used for forecasting tasks, label extractors have access to all loaded data points, irrespective of their timestamp.
Edges, feature matrices, and labels of all snapshots are bundled in a \texttt{NetworkSnapshots} object, which implementations of \texttt{NetworkExporter} convert into network objects keyed by timestamp. 
\texttt{Git2KG} ships exporters for PyTorch Geometric~\cite{DBLP:journals/corr/abs-1903-02428}, PathPyG\footnote{https://www.pathpy.net/}, and NetworkX\footnote{https://networkx.org}. 
\begin{figure}
    \centering
    \includegraphics[width=.9\linewidth]{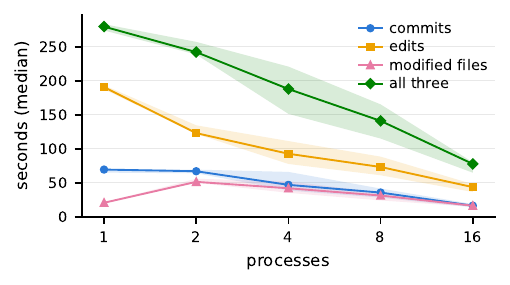}
    \caption{Loading time of the Git-based extraction stages with different numbers of cores in seconds. The line represents the median loading times across five runs. The band represents the minimum and maximum loading time of five runs. The runs were conducted on an AMD Ryzen AI 9 HX 370.}\label{fig:performance}
\end{figure}

\section{Evaluation and Demonstration}\label{sec:evaluation}
   
In this section, we present evaluation results regarding Git mining performance and actor disambiguation. 
We further demonstrate potential use cases for \acp{tkg} mined with \texttt{Git2KG}. 
For the analysis and demonstration, we use the \ac{tkg} mined from the \texttt{xz-utils} project\footnote{https://github.com/tukaani-project/xz}. 
The xz-utils project is a widely used compression library which gained massive attention in March 2024, when a malicious backdoor was discovered in a published tarball of the project. 
We mine the repository with \texttt{Git2KG} and extract \ac{kg} snapshots from monthly time windows. 
The dataset contains $3142$ commits, $447650$ edits, $252$ issues with $1131$ comments, $236$ issue events, and $244$ reviews.
We only use snapshots after October 2022, as the project started using the issue tracker that month. 

\begin{figure}
    \centering
    \includegraphics[width=.9\linewidth]{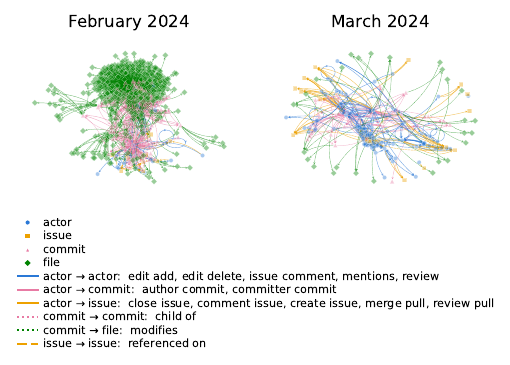}
    \caption{Visualization of two consecutive monthly snapshots of the \ac{tkg} mined from the repository of xz-utils. Color and shape of the nodes indicate the entity type. The color of the edge refers to the entity type of its end node. The shape of the edge's line indicates the entity type of its start node. The security breach was discovered in March 2024.}\label{fig:xz-snapshots}
\end{figure}
In Fig.~\ref{fig:xz-snapshots}, we visualize two snapshots from the \ac{tkg} using graph exports from the \texttt{NetworkXExporter} of \texttt{Git2KG}.    
For February 2024, we observe that the graph is dominated by relations and entities related to commits, indicating strong development activity in that time slice. 
We note that this is not necessarily related to the introduction of the backdoor, but might be just a consequence of planned releases in that month. 
In March 2024, when the backdoor was discovered, we see far more relations encoding communication on the issue tracker, which is likely due to users commenting on the publication and related posts. 

\subsection{Mining Performance}

Fig.~\ref{fig:performance} shows the median loading times in seconds for the Git-related data points on the xz-utils repository across five runs. 
Comparing the loading time using only one process with the loading time using multiple processes, we observe that the median loading time for all data points drops from $279.56$s to $77.59$s when using $16$ processes instead of one, i.e., the loading is roughly $3.6$ times faster. 
We observe the strongest speedup of $4.38$ for the edit data points, which is expected since loading edits is computationally the most expensive operation. 
For loading modified files, we see a weak speedup of $1.27$, comparing $16$ processes with one. 
Interestingly, the modified files are actually loaded more slowly for two, four, and eight processes than just using one process. 
This could be due to the overhead of chunking the commits and passing them to the subprocesses. 
Since the loading of a modified file data point is essentially equal to converting PyDriller's \texttt{ModifiedFile}, the operation is computationally cheap, and the orchestrating overhead for multiprocessing has a bigger impact. 
\begin{figure}[!t]
    \centering
    \includegraphics[width=.9\linewidth]{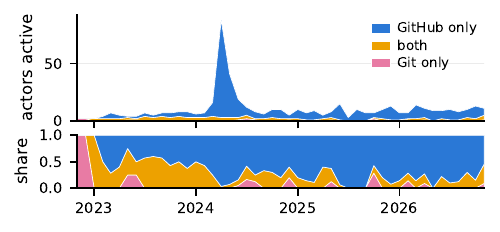}
    \caption{Total number and share of actor entities that appear as the start node of an edge across snapshots mined from the xz-utils repository. 
    The color of the curves indicates whether the entity was subject to an edge mined from Git, GitHub, or both. 
    }\label{fig:alignment}
\end{figure}

\subsection{Actor Alignment}
If Git authors and committers were mapped to different entities than their GitHub user accounts, the two subgraphs would be disconnected.
Fig.~\ref{fig:alignment} shows, for each monthly snapshot, the number and share of actors that are subject to edges from Git only, GitHub only, or both. 
We observe that in most snapshots, actors that are active in Git are also active on GitHub, as shown by the orange curve. 
This demonstrates that actor alignment connects the two subgraphs.  
The peak of GitHub-only actors around March 2024 likely reflects the increased communication following the disclosure of the backdoor.

\section{Conclusion and Limitations}\label{sec:conclusion} 
We present \texttt{Git2KG}, a \ac{tkg} mining library for GitHub repositories, that aligns entities from Git and GitHub to form connected \ac{kg} snapshots.
The library is extendable in terms of feature and label computation, data export, and loading persistence, and therefore enables users to adjust the setup to their needs and integrate the library into existing ecosystems.
We mine \acp{kg} with a schema of relations that describe how actors interact with each other, with the issue tracker, and with repository artifacts. 
However, we do not represent source code content or structure, continuous integration outcomes, forks, or line-level code review comments. 
Our models focus on collaborative patterns and are currently not suitable for analysis and learning tasks on program semantics. 
These limitations are subject of our further research.

While we mine data from Git logs offline, we still depend on APIs of GitHub to load the corresponding data points. 
Therefore, we rely on a volatile data source which might change its interface in the future. 
We further note that, while our tool is applicable to every repository on GitHub, users still need to carefully select the repositories they mine~\cite{DBLP:conf/msr/KalliamvakouGBSGD14}. 
Migrations from other code hosting tools to GitHub or the usage of external issue trackers might obfuscate the mined data. 
In future research, our tool could be used for large-scale analyses of the open-source ecosystem, similar to~\cite{DBLP:conf/icse/GoteMSS22}, or machine learning tasks, like predicting developer turnover~\cite{DBLP:conf/sigsoft/FoucaultPBMF15}. 

\bibliographystyle{IEEEtran}
\bibliography{IEEEabrv, references}

\end{document}